# An Alternative Derivation of the Landau–Lifshitz–Gilbert Equation for Saturated Ferromagnets

Jiashi Yang
Department of Mechanical and Materials Engineering
University of Nebraska-Lincoln, Lincoln, NE 68588-0526, USA
jyang1@unl.edu

**Abstract**
The Landau–Lifshitz–Gilbert equation for rigid and saturated ferromagnets is derived using a two-continuum model constructed by H.F. Tiersten for elastic and saturated ferromagnets. The relevant basic laws of physics are applied systematically to the two continua or their combination. The exchange interaction is introduced into the model through surface distributed magnetic couples. This leads to a continuum theory with magnetization gradients in the stored energy density. The saturation condition of the magnetization functions as constraints on the energy density and has implications in the constitutive relations.

## 1. Introduction

In ferromagnetic solids, microscopic magnetic moments may align themselves to form regions with spontaneous magnetization below the Curie temperature. This is because of the exchange interaction among neighboring magnetic moments, which is quantum mechanical in nature. In saturated ferromagnets, the magnetization vector has a fixed magnitude and can change its direction only. A disturbance of the aligned magnetic moments can propagate as what is called spin waves which are governed by the well-known Landau–Lifshitz (LL) equation. The LL equation can be obtained from continuum modeling by including the magnetization gradient into the stored energy density [1], or from discrete models of interacting magnetic moments followed by taking the continuum limit [2]. When damping is included, the LL equation becomes the more general Landau–Lifshitz–Gilbert (LLG) equation [3]. There are also relatively recent reports on the derivation of the LL or LLG equation [4].

In deformable materials, spin waves may interact with elastic waves which is referred to as phonon-magnon interaction. Continuum theories of phonon-magnon interactions in saturated ferromagnetoelastic solids can be found in [5-13]. In particular, H.F. Tiersten constructed a two-continuum model from which the theory of saturated ferromagnetoelasticity was established by applying the relevant physical laws to the two continua or their combination [5]. While the theory of rigid ferromagnets may be viewed as a special case of the theory of elastic ferromagnets, the reduction of the theory of elastic ferromagnets in [5] to rigid ferromagnets needs some effort because the theory in [5] is rather complicated. Therefore it is worthwhile to construct the theory of rigid ferromagnets directly from the two-continuum model in [5] which is carried out below. Different from [5] where the magnetic field vector **H** is used to express the magnetic couple on a magnetic moment, the magnetic induction vector **B** is used below which is the same as what is in the later paper [9] by H.F. Tiersten. The reason of using **B** instead of **H** is because of the current-loop model of magnetic moments in [9,14] where the couple on a current loop is calculated from the Biot–Savart law in terms of **B**. Another difference from [5] is that the SI unit system is used below instead of Gaussian units.

## 2. Saturated Ferromagnets

We use the Cartesian tensor notation [15]. The spatial coordinates are written as $x_k$ or **x**. $V$, $S$ and $C$ represent volumes, surfaces and curves fixed in space. The outward unit normal of $S$ is **n**. Consider the following vector field of the magnetization density per unit volume in a stationary and rigid ferromagnet:

$$\mathbf{M} = \mathbf{M}(\mathbf{x}, t). \qquad (1)$$

For saturated ferromagnets we have the following saturation condition:

$$\mathbf{M} \cdot \mathbf{M} = M_s^2, \qquad (2)$$

where $M_s$ is a constant (saturation magnetization) and $s$ is not a tensor index. Mathematically, (2) is a constraint on $\mathbf{M}$. With differentiations with respect to $t$ and/or $\mathbf{x}$, (2) implies that

$$M_k \frac{\partial M_k}{\partial t} = 0,$$
$$M_k M_{k,l} = 0, \qquad (3)$$
$$M_{k,l} \frac{\partial M_k}{\partial t} + M_k \frac{\partial M_{k,l}}{\partial t} = 0.$$

Although the magnitude of $\mathbf{M}$ cannot change because of the saturation condition, $\mathbf{M}$ can still change its direction as described by the rotation or angular displacement $\delta\theta = |\delta\mathbf{M}|/|\mathbf{M}|$ in Fig. 1.

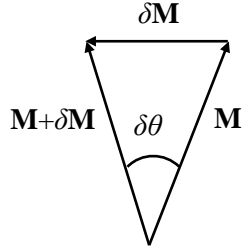

Fig. 1. Change of direction of $\mathbf{M}$.

We introduce an angular displacement vector $\boldsymbol{\delta\theta}$ by

$$\boldsymbol{\delta\theta} = \delta\theta \frac{\mathbf{M}}{|\mathbf{M}|} \times \frac{\delta\mathbf{M}}{|\delta\mathbf{M}|} = \frac{|\delta\mathbf{M}|}{|\mathbf{M}|} \frac{\mathbf{M}}{|\mathbf{M}|} \times \frac{\delta\mathbf{M}}{|\delta\mathbf{M}|} = \frac{1}{M_s^2} \mathbf{M} \times (\delta\mathbf{M}). \qquad (4)$$

Then

$$\boldsymbol{\delta\theta} \times \mathbf{M} = \frac{1}{M_s^2}(\mathbf{M} \times \delta\mathbf{M}) \times \mathbf{M} = \frac{1}{M_s^2}\left[(\mathbf{M} \cdot \mathbf{M})\delta\mathbf{M} - (\mathbf{M} \cdot \delta\mathbf{M})\mathbf{M}\right] = \delta\mathbf{M}, \qquad (5)$$

where the following vector identity has been used [16]:

$$\mathbf{A} \times (\mathbf{B} \times \mathbf{C}) = (\mathbf{A} \cdot \mathbf{C})\mathbf{B} - (\mathbf{A} \cdot \mathbf{B})\mathbf{C}. \qquad (6)$$

We also introduce an angular velocity vector for a saturated $\mathbf{M}$ through

$$\boldsymbol{\omega} = \lim_{\delta t \to 0} \frac{\boldsymbol{\delta\theta}}{\delta t} = \frac{1}{M_s^2} \mathbf{M} \times \frac{\partial \mathbf{M}}{\partial t}. \qquad (7)$$

The power of a magnetic couple $\boldsymbol{\Gamma} = \mathbf{M} \times \mathbf{B}$ on a saturated $\mathbf{M}$ during an angular motion of $\mathbf{M}$ is given by

$$w^M = \boldsymbol{\Gamma} \cdot \boldsymbol{\omega} = (\mathbf{M} \times \mathbf{B}) \cdot \left(\frac{1}{M_s^2} \mathbf{M} \times \frac{\partial \mathbf{M}}{\partial t}\right)$$
$$= \frac{1}{M_s^2}\left[(\mathbf{M} \cdot \mathbf{M})\left(\mathbf{B} \cdot \frac{\partial \mathbf{M}}{\partial t}\right) - \left(\mathbf{M} \cdot \frac{\partial \mathbf{M}}{\partial t}\right)(\mathbf{B} \cdot \mathbf{M})\right] = \mathbf{B} \cdot \frac{\partial \mathbf{M}}{\partial t}, \qquad (8)$$

where we have used the following vector identity [16]:

$$(\mathbf{u} \times \mathbf{v}) \cdot (\mathbf{w} \times \mathbf{t}) = (\mathbf{u} \cdot \mathbf{w})(\mathbf{v} \cdot \mathbf{t}) - (\mathbf{u} \cdot \mathbf{t})(\mathbf{v} \cdot \mathbf{w}). \qquad (9)$$

Following [5], we write the angular momentum of $\mathbf{M}$ as

$$\mathbf{L} = \frac{1}{\gamma}\mathbf{M}, \qquad (10)$$

where $\gamma$ is the gyromagnetic ratio which is a negative number.

### 3. Two-Continuum Model

Saturated ferromagnetic insulators can be modeled by the two interpenetrating and interacting continua shown in Fig. 2 [5]. One is called the spin continuum which carries distributed magnetic moments. The other is the lattice continuum which is assumed to be rigid. The two continua cannot displace relatively with respect to each other, but the magnetic moments can rotate with respect to the lattice. The two continua interact through a local force $\mathbf{f}^L$ and a local couple $\mathbf{c}^L$ produced by an effective local magnetic induction $\mathbf{B}^L$ [9,14], i.e.,

$$\mathbf{c}^L = \mathbf{M} \times \mathbf{B}^L. \tag{11}$$

As a moment vector, $\mathbf{c}^L$ is shown by double arrows in Fig. 2. The spin continuum is assumed to be massless. It experiences a magnetic body force $\mathbf{f}^M$ and a magnetic body couple $\mathbf{c}^M$ produced by the Maxwellian magnetic induction $\mathbf{B}^M$:

$$\mathbf{c}^M = \mathbf{M} \times \mathbf{B}^M. \tag{12}$$

The spin continuum also experiences a distributed couple $\mathbf{M} \times \mathbf{F}$ per unit area on its boundary surface due to an effective exchange field $\mathbf{F}$ whose nature is quantum mechanical. The use of a surface distribution of $\mathbf{F}$ for the short-range exchange interaction is the key of the formulation in [5]. Since $\mathbf{F}$ and $\mathbf{B}^L$ act on $\mathbf{M}$ through cross products, the components of $\mathbf{F}$ and $\mathbf{B}^L$ along $\mathbf{M}$ have no contributions. Hence it can be assumed that [5]

$$\mathbf{F} \cdot \mathbf{M} = 0, \tag{13}$$

$$\mathbf{B}^L \cdot \mathbf{M} = 0. \tag{14}$$

The lattice continuum is under the usual mechanical surface traction $\mathbf{t}$ and mechanical body force $\mathbf{f}$, in addition to the interactions with the spin continuum through $-\mathbf{f}^L$ and $-\mathbf{c}^L$.

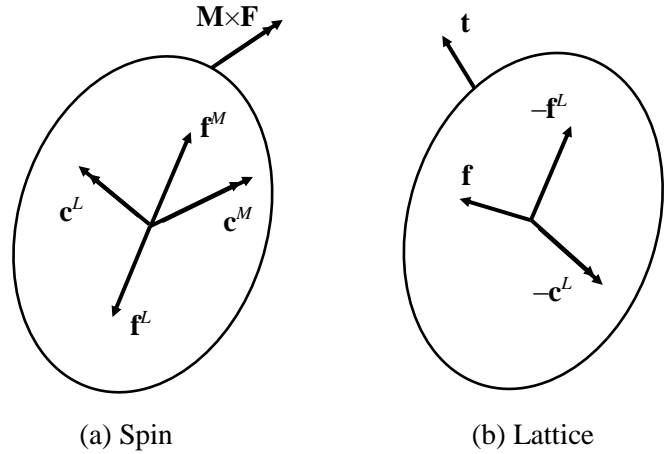

(a) Spin    (b) Lattice

Fig. 2. Separate spin and lattice continua.

The combined continuum in Fig. 3 shows the loads external to the two continua only without their interactions which are internal.

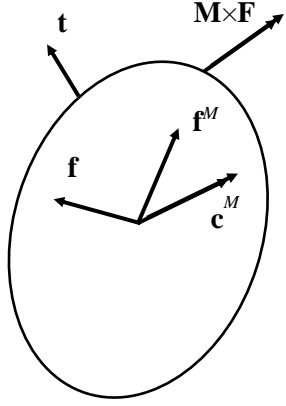

Fig. 3. Combined continuum of lattice and spin.

## 4. Integral Balance Laws

The relevant balance laws from physics are

$$\int_S \mathbf{n} \cdot \mathbf{B}^M dS = 0, \tag{15}$$

$$\oint_C \mathbf{H} \cdot d\mathbf{x} = 0, \tag{16}$$

$$\int_V (\mathbf{f}^M + \mathbf{f}^L) dV = 0, \tag{17}$$

$$\frac{\partial}{\partial t} \int_V \frac{\mathbf{M}}{\gamma} dV = \int_S \mathbf{M} \times \mathbf{F} dS + \int_V \mathbf{M} \times \left(\mathbf{B}^M + \mathbf{B}^L\right) dV, \tag{18}$$

$$\frac{\partial}{\partial t} \int_V U dV = \int_S \left(\mathbf{F} \cdot \frac{\partial \mathbf{M}}{\partial t} - \mathbf{n} \cdot \mathbf{q}\right) dS + \int_V \left(-\mathbf{M} \cdot \frac{\partial \mathbf{B}^M}{\partial t} + R\right) dV, \tag{19}$$

$$\frac{\partial}{\partial t} \int_V \eta dV \geq \int_V \frac{R}{\theta} dV - \int_S \frac{\mathbf{q} \cdot \mathbf{n}}{\theta} dS, \tag{20}$$

where $U$ is the internal energy density per unit volume, $R$ the body heat source per unit volume, $\mathbf{q}$ the heat flux vector, $\theta$ the absolute temperature and $\eta$ the entropy density per unit volume. (17) and (18) are the linear and angular momentum equations for the spin continuum alone. Those for the lattice continuum alone are not of interest and therefore are left out. (19) and (20) are the energy equation and the second law of thermodynamics of the combined continuum. The magnetic power in (19), $-\mathbf{M} \cdot \partial \mathbf{B}^M/\partial t$ [9,14], is based on the current-loop model for magnetic moments. This is the same as [9] but is different from [5] where (8) was used for both of the two terms for magnetic powers on the right-hand side of (19). These two different approaches lead to theories with the same mathematical structure, but the physical interpretations of the energy densities involved are different.

## 5. Differential Balance Laws

The differential forms of (15) and (16) for the quasistatic magnetic fields are

$$\nabla \cdot \mathbf{B}^M = 0, \tag{21}$$
$$\nabla \times \mathbf{H} = 0. \tag{22}$$

Since the spin continuum is massless, the linear momentum equation in (17) simply leads to

$$\mathbf{f}^M + \mathbf{f}^L = 0. \tag{23}$$

(23) is not needed in the theoretical framework below. The linear momentum equation of the spin continuum will be necessary when the spin continuum is no longer massless and can displace from the lattice continuum. For the angular momentum equation in (18), we introduce an exchange tensor $\mathbf{A}$ by [5]

$$\mathbf{F} = -\mathbf{n} \cdot \mathbf{A}, \tag{24}$$

which is restricted by (13) through

$$\mathbf{A} \cdot \mathbf{M} = 0. \tag{25}$$

Then the angular momentum equation in (18) can be brought into the following differential form using (24) and the divergence theorem:

$$\varepsilon_{ijk} M_j \left( -A_{lk,l} + B_k^M + B_k^L \right) - \varepsilon_{ijk} A_{lk} M_{j,l} = \frac{1}{\gamma} \frac{\partial M_i}{\partial t}. \tag{26}$$

As a cross product, the first term on the left-hand side of (26) is perpendicular to $\mathbf{M}$. Dotting both sides of (26) by $\mathbf{M}$, we have

$$-M_i \varepsilon_{ijk} A_{lk} M_{j,l} = \frac{1}{\gamma} M_i \frac{\partial M_i}{\partial t}. \tag{27}$$

The right-hand side of (27) vanishes because of the constraint due to saturation in (3)$_1$. Then (27) reduces to

$$-M_i \varepsilon_{ijk} A_{lk} M_{j,l} = 0. \tag{28}$$

To satisfy (28), we impose the following restriction on $\mathbf{A}$ [5]:

$$A_{lk} M_{j,l} = A_{lj} M_{k,l}. \tag{29}$$

With the use of (29), the angular momentum equation in (26) reduces to

$$\frac{1}{\gamma} \frac{\partial M_i}{\partial t} = \varepsilon_{ijk} M_j \left( -A_{lk,l} + B_k^M + B_k^L \right). \tag{30}$$

(30) can be written as

$$\frac{1}{\gamma} \frac{\partial \mathbf{M}}{\partial t} = \mathbf{M} \times \mathbf{B}^{eff}, \tag{31}$$

where

$$\begin{aligned} \mathbf{B}^{eff} &= \mathbf{B}^{ex} + \mathbf{B}^M + \mathbf{B}^L, \\ B_k^{ex} &= -A_{lk,l}. \end{aligned} \tag{32}$$

(32) is the well-known LL equation in the literature. $\mathbf{B}^{eff}$ is the total effective magnetic induction. $\mathbf{B}^{ex}$ is the effective exchange induction. If we take a dot product of both sides of the angular momentum equation in (30) with the following vector which is along the $\boldsymbol{\omega}$ in (7):

$$\varepsilon_{imn} M_m \frac{\partial M_n}{\partial t}, \tag{33}$$

we obtain

$$\frac{\partial M_k}{\partial t} \left( -A_{lk,l} + B_k^M + B_k^L \right) = 0, \tag{34}$$

which will be useful later. The differential forms of (19) and (20) can be obtained using the divergence theorem as:

$$\frac{\partial U}{\partial t} = -A_{ij,i} \left( \frac{\partial M_j}{\partial t} \right) - A_{ij} \left( \frac{\partial M_j}{\partial t} \right)_{,i} - M_j \frac{\partial B_j^M}{\partial t} + R - q_{i,i}, \tag{35}$$

$$\frac{\partial \eta}{\partial t} \geq \frac{R}{\theta} - \left( \frac{q_i}{\theta} \right)_{,i}. \tag{36}$$

## 6. Constitutive Relations

With the use of (34), the energy equation in (35) becomes

$$\frac{\partial U}{\partial t} = -B_j^M \frac{\partial M_j}{\partial t} - B_j^L \frac{\partial M_j}{\partial t} - A_{ij} \left( \frac{\partial M_j}{\partial t} \right)_{,i} - M_j \frac{\partial B_j^M}{\partial t} + R - q_{i,i}. \tag{37}$$

Under the following Legendre transform:
$$F = U + B_i^M M_i - \theta\eta,\tag{38}$$
(37) becomes
$$\frac{\partial F}{\partial t} + \frac{\partial \theta}{\partial t}\eta + \theta\frac{\partial \eta}{\partial t} = -A_{ij}\left(\frac{\partial M_j}{\partial t}\right)_{,i} - B_j^L \frac{\partial M_j}{\partial t} + R - q_{i,i}.\tag{39}$$

Substituting (38) into (36) and eliminating $R$ using (39), we obtain the Clausius–Duhem inequality as
$$-\left(\frac{\partial F}{\partial t} + \frac{\partial \theta}{\partial t}\eta\right) - A_{ij}\left(\frac{\partial M_j}{\partial t}\right)_{,i} - B_j^L \frac{\partial M_j}{\partial t} - \frac{q_i}{\theta}\theta_{,i} \geq 0.\tag{40}$$

For constitutive relations we break $\mathbf{B}^L$ into reversible and dissipative parts and assume
$$\mathbf{B}^L = {}^R\mathbf{B}^L(M_i; M_{j,i}; \theta) + {}^D\mathbf{B}^L(M_i; M_{j,i}; \theta; \theta_{,j}; \dot{M}_i),$$
$$\mathbf{q} = \mathbf{q}(M_i; M_{j,i}; \theta; \theta_{,j}; \dot{M}_i).\tag{41}$$

For simplicity we are not considering possible dissipations related to the exchange interaction. The reversible part of $(41)_1$ is chosen to satisfy
$$\frac{\partial F}{\partial t} = -{}^R B_j^L \frac{\partial M_j}{\partial t} - A_{ij}\left(\frac{\partial M_j}{\partial t}\right)_{,i} - \eta\frac{\partial \theta}{\partial t}.\tag{42}$$

Then the energy equation in (39) and the Clausius–Duhem inequality in (40) become
$$\theta\frac{\partial \eta}{\partial t} = -{}^D B_j^L \frac{\partial M_j}{\partial t} + R - q_{i,i},\tag{43}$$
$$-{}^D B_j^L \frac{\partial M_j}{\partial t} - \frac{q_i}{\theta}\theta_{,i} \geq 0.\tag{44}$$

(43) is the heat or dissipation equation. With
$$\left(\frac{\partial M_j}{\partial t}\right)_{,i} = \frac{\partial}{\partial t}(M_{j,i}),\tag{45}$$
we write (42) as
$$\frac{\partial F}{\partial t} = -{}^R B_j^L \frac{\partial M_j}{\partial t} - A_{ij}\frac{\partial}{\partial t}(M_{j,i}) - \eta\frac{\partial \theta}{\partial t}.\tag{46}$$

Let
$$F = F(M_i; M_{j,i}; \theta).\tag{47}$$

Then
$$\frac{\partial F}{\partial t} = \frac{\partial F}{\partial M_i}\frac{\partial M_i}{\partial t} + \frac{\partial F}{\partial(M_{j,i})}\frac{\partial}{\partial t}(M_{j,i}) + \frac{\partial F}{\partial \theta}\frac{\partial \theta}{\partial t}.\tag{48}$$

We substitute (48) into (46) and use Lagrange multipliers $\lambda$ and $L_i$ to introduce the constrains in $(3)_{1,3}$. This yields
$$\frac{\partial F}{\partial M_i}\frac{\partial M_i}{\partial t} + \frac{\partial F}{\partial(M_{j,i})}\frac{\partial}{\partial t}(M_{j,i}) + \frac{\partial F}{\partial \theta}\frac{\partial \theta}{\partial t}$$
$$= -A_{ij}\frac{\partial}{\partial t}(M_{j,i}) - {}^R B_j^L \frac{\partial M_j}{\partial t} - \eta\frac{\partial \theta}{\partial t} + \lambda M_i \frac{\partial M_i}{\partial t} + L_i\left[M_{j,i}\frac{\partial M_j}{\partial t} + M_j\frac{\partial}{\partial t}(M_{j,i})\right],\tag{49}$$
or
$$-\left(\frac{\partial F}{\partial \theta} + \eta\right)\frac{\partial \theta}{\partial t} - \left({}^R B_i^L - \lambda M_i - L_j M_{i,j} + \frac{\partial F}{\partial M_i}\right)\frac{\partial M_i}{\partial t} - \left[A_{ij} - L_i M_j + \frac{\partial F}{\partial(M_{j,i})}\right]\frac{\partial}{\partial t}(M_{j,i}) = 0.\tag{50}$$

(50) implies the following reversible constitutive relations:

$$^R B_i^L = -\frac{\partial F}{\partial M_i} + \lambda M_i + L_m M_{i,m},$$

$$A_{ij} = -\frac{\partial F}{\partial (M_{j,i})} + L_i M_j, \quad (51)$$

$$\eta = -\frac{\partial F}{\partial \theta}.$$

Since we have assumed (13) and (14) which impose restrictions on **F** and **B**$^L$, we can use (13) and (14) to determine $L_i$ and $\lambda$ [5]. From (25) which is an implication of (13), we have $\mathbf{A} \cdot \mathbf{M} = 0$ or, with the use of (51)$_2$,

$$A_{ij} M_j = -\frac{\partial F}{\partial (M_{j,i})} M_j + L_i M_j M_j = 0. \quad (52)$$

(52) determines that

$$L_i = \frac{1}{M_s^2} \frac{\partial F}{\partial (M_{k,i})} M_k. \quad (53)$$

From (14) we have

$$\mathbf{B}^L \cdot \mathbf{M} = {}^R\mathbf{B}^L \cdot \mathbf{M} + {}^D\mathbf{B}^L \cdot \mathbf{M} = 0. \quad (54)$$

As a sufficient condition of (54), we impose

$$^R\mathbf{B}^L \cdot \mathbf{M} = 0, \quad {}^D\mathbf{B}^L \cdot \mathbf{M} = 0. \quad (55)$$

With the use of (51)$_1$, we have, from (55)$_1$,

$$^R B_i^L M_i = -\frac{\partial F}{\partial M_i} M_i + \lambda M_i M_i + L_j M_i M_{i,j} = -\frac{\partial F}{\partial M_i} M_i + \lambda M_i M_i = 0, \quad (56)$$

where (3)$_2$ has been used. (56) determines that

$$\lambda = \frac{1}{M_s^2} \frac{\partial F}{\partial M_k} M_k. \quad (57)$$

Using (53) and (57), we write (51)$_{1,2}$ as

$$^R B_i^L = -\frac{\partial F}{\partial M_i} + \frac{1}{M_s^2} \frac{\partial F}{\partial M_k} M_k M_i + \frac{1}{M_s^2} \frac{\partial F}{\partial (M_{k,j})} M_k M_{i,j},$$

$$A_{ij} = -\frac{\partial F}{\partial (M_{j,i})} + \frac{1}{M_s^2} \frac{\partial F}{\partial (M_{k,i})} M_k M_j, \quad (58)$$

where **A** is restricted by (29). The dissipative constitutive relations are restricted by (44) and (55)$_2$.

When damping is present, the LLG equation in the literature has the following from:

$$\frac{1}{\gamma} \frac{\partial \mathbf{M}}{\partial t} = \mathbf{M} \times \mathbf{B}^{eff} - \beta \mathbf{M} \times \frac{\partial \mathbf{M}}{\partial t}, \quad (59)$$

where $\beta$ is a damping coefficient. (59) can be written as

$$\frac{1}{\gamma} \frac{\partial \mathbf{M}}{\partial t} = \mathbf{M} \times \left( \mathbf{B}^{eff} - \beta \frac{\partial \mathbf{M}}{\partial t} \right). \quad (60)$$

On the other hand, with the theoretical framework developed in the above, from (31), (32) and (41), we can write

$$\frac{1}{\gamma} \frac{\partial \mathbf{M}}{\partial t} = \mathbf{M} \times (\mathbf{B}^{ex} + \mathbf{B}^M + {}^R\mathbf{B}^L + {}^D\mathbf{B}^L). \quad (61)$$

Comparing (59) and (61), we identify

$$^D\mathbf{B}^L = -\beta \frac{\partial \mathbf{M}}{\partial t}, \quad (62)$$

which satisfies (55)$_2$. Thus the theoretical framework derived in the above can reduce to the LLG equation. If the thermal term in (44) is neglected, (44) reduces to

$$-{}^D\mathbf{B}^L \cdot \frac{\partial \mathbf{M}}{\partial t} \geq 0. \tag{63}$$

Then, with the use of (62), we have

$$\beta \frac{\partial \mathbf{M}}{\partial t} \cdot \frac{\partial \mathbf{M}}{\partial t} \geq 0, \tag{64}$$

which implies that $\beta \geq 0$.

## 7. Summary of Equations

In summary, the field equations are

$$\nabla \cdot \mathbf{B}^M = 0, \tag{65}$$

$$\nabla \times \mathbf{H} = 0, \tag{66}$$

$$\frac{1}{\gamma} \frac{\partial M_i}{\partial t} = \varepsilon_{ijk} M_j \left( -A_{lk,l} + B_k^M + B_k^L \right), \tag{67}$$

$$\theta \frac{\partial \eta}{\partial t} = -{}^D B_j^L \frac{\partial M_j}{\partial t} + R - q_{i,i}. \tag{68}$$

For constitutive relations, we have

$$\mathbf{B}^L = {}^R\mathbf{B}^L + {}^D\mathbf{B}^L \left( M_i; M_{j,i}; \theta; \theta_{,j}; \dot{M}_i \right),$$

$$\mathbf{q} = \mathbf{q}\left( M_i; M_{j,i}; \theta; \theta_{,j}; \dot{M}_i \right), \tag{69}$$

$$F = F\left( M_i; M_{j,i}; \theta \right), \tag{70}$$

$${}^R B_i^L = -\frac{\partial F}{\partial M_i} + \frac{1}{M_s^2} \frac{\partial F}{\partial M_k} M_k M_i + \frac{1}{M_s^2} \frac{\partial F}{\partial (M_{k,j})} M_k M_{i,j},$$

$$A_{ij} = -\frac{\partial F}{\partial (M_{j,i})} + \frac{1}{M_s^2} \frac{\partial F}{\partial (M_{k,i})} M_k M_j, \tag{77}$$

$$\eta = -\frac{\partial F}{\partial \theta},$$

which are restricted by (29), (44) and (55)$_2$. (66) allows the introduction of a scalar potential $\psi$ as follows:

$$\mathbf{H} = -\nabla \psi, \tag{72}$$

In addition, we have

$$\mathbf{B}^M = \mu_0 (\mathbf{H} + \mathbf{M}). \tag{73}$$

Then (65), (67) and (68) can be written as five equations for $\psi$, $\theta$ and the three components of **M**. On a boundary surface with an outward unit normal **n**, possible boundary conditions are the prescriptions of [5]

$$\psi \quad \text{or} \quad \mathbf{n} \cdot \mathbf{B}^M,$$

$$\theta \quad \text{or} \quad \mathbf{n} \cdot \mathbf{q}, \tag{74}$$

$$\delta\boldsymbol{\theta} \quad \text{or} \quad \mathbf{n} \cdot \mathbf{A} \times \mathbf{M},$$

where $\theta$ is the absolute temperature and $\delta\boldsymbol{\theta}$ is the angular displacement of **M**. In dynamic problems, instead of $\delta\boldsymbol{\theta}$, its time derivative $\boldsymbol{\omega}$ may be prescribed.

## 8. Cubic Crystals

As a simple example of constitutive relations, consider the case of small magnetization with small magnetization gradients. We take

$$F = \frac{1}{2}\chi_{mn}M_m M_n + \frac{1}{2}\alpha_{mn}M_{k,m}M_{k,n}, \quad \chi_{mn} = \chi_{nm}, \quad \alpha_{mn} = \alpha_{nm}, \tag{76}$$

which consists of two of the many possible terms in a long expression in [5]. The symmetry of $\alpha_{mn}$ ensures that (29) is satisfied. We consider the special case of cubic crystals of class (m3m) with $\alpha_{im} = \alpha\delta_{im}$. Then (71)$_2$ generates

$$A_{ij} = -\alpha\delta_{im}M_{j,m} = -\alpha M_{j,i}. \tag{77}$$

The effective exchange field $\mathbf{B}^{ex}$ has the following expression according to (32):

$$B_j^{ex} = -A_{ij,i} = \alpha M_{j,ii}, \quad \mathbf{B}^{ex} = \alpha\nabla^2 \mathbf{M}, \tag{78}$$

which is a common expression for $\mathbf{B}^{ex}$ (or $\mathbf{H}^{ex}$) in the literature for cubic crystals, e.g., [4] where an expression like (78) is obtained for $\mathbf{H}^{ex}$ from a discrete model of magnetic moments in cubic crystals.

## 9. Conclusions

Equations for rigid and saturated ferromagnets are derived systematically from the two-continuum model constructed by H.F. Tiersten. The constitutive relations satisfy the saturation condition of magnetization. It is a magnetization gradient theory in the sense that the stored energy depends on the magnetization gradient in addition the magnetization itself. The equations derived can reduce to the LLG equation as a special case. The constitutive equations are for general material anisotropy and can be specialized to any crystal classes.

## References

[1] L.D. Landau and E.M. Lifshitz, *Electrodynamics of Continuous Media*, 2nd ed., Butterworth–Heinemann, Linacre House, Jordan Hill, Oxford, 1984.

[2] E.Y. Tsymbal, *Introduction to Solid State Physics*, https://unlcms.unl.edu/cas/physics/tsymbal/teaching/SSP-927/index.shtml.

[3] T.L. Gilbert, A phenomenological theory of damping in ferromagnetic materials, *IEEE Trans. Magn.*, 40, 3443–3449, 2004.

[4] M. Krawczyk, M.L. Sokolovskyy, J.W. Klos and S. Mamica, On the formulation of the exchange field in the Landau–Lifshitz equation for spin-wave calculation in magnonic crystals, *Adv. Condens. Matter Phys.*, 2012, 764783, 2012.

[5] H.F. Tiersten, Coupled magnetomechanical equation for magnetically saturated insulators, *J. Math. Phys.*, 5, 1298–1318, 1964.

[6] H.F. Tiersten, Variational principle for saturated magnetoelastic insulators, *J. Math. Phys.*, 6, 779–787, 1965.

[7] H.F. Tiersten, Thickness vibrations of saturated magnetoelastic plates, *J. Appl. Phys.*, 36, 2250–2259, 1965.

[8] H.F. Tiersten, Surface coupling in magnetoelastic interactions, in: *Surface Mechanics*, ASME, New York, 1969, pp.120–142.

[9] H.F. Tiersten and C.F. Tsai, On the interaction of the electromagnetic field with heat conducting deformable insulators, *J. Math. Phys.*, 13, 361–378, 1972.

[10] W.F. Brown Jr., Theory of magnetoelastic effects in ferromagnetism, *J. Appl. Phys.*, 36, 994–1000, 1965.

[11] W.F. Brown Jr., *Magnetoelastic Interactions*, Springer–Verlag, New York, 1966.

[12] G.A. Maugin and A.C. Eringen, Deformable magnetically saturated media. I. Field equations, *J. Math. Phys.* 13, 143–155, 1972.

[13] G.A. Maugin and A.C. Eringen, Deformable magnetically saturated media. II. Constitutive theory, *J. Math. Phys.* 13, 1334–1347, 1972.

[14] H.F. Tiersten, *A Development of the Equations of Electromagnetism in Material Continua*, Springer, New York, 1990.

[15] A.C. Eringen, *Mechanics of Continua*, Robert E. Krieger, Huntington, New York, 1980.

[16] G.E. Hay, *Vector and Tensor Analysis*, Dover, New York, 1958.